\documentclass[acmtrecs]{acmtrans2m} 
\pdfoutput=1

\usepackage{epsfig}

\title{Searching for Transient Pulses with the ETA Radio Telescope}

\author{C.D.~PATTERSON, S.W.~ELLINGSON, B.S.~MARTIN and K.~DESHPANDE \\
Department of Electrical and Computer Engineering \\
Virginia Tech
\and
J.H.~SIMONETTI, M.~KAVIC and S.E.~CUTCHIN \\
Department of Physics \\
Virginia Tech
}

\begin{abstract}
Array-based, direct-sampling radio telescopes have computational and communication requirements unsuited to conventional computer and cluster architectures.
Synchronization must be strictly maintained across a large number of parallel data streams, from A/D conversion, through operations such as beamforming, to dataset recording.
FPGAs supporting multi-gigabit serial I/O are ideally suited to this application.
We describe a recently-constructed radio telescope called ETA having all-sky observing capability for detecting low frequency pulses from transient events such as gamma ray bursts and primordial black hole explosions.
Signals from 24 dipole antennas are processed by a tiered arrangement of 28 commercial FPGA boards  and 4 PCs with FPGA-based data acquisition cards, connected with custom I/O adapter boards supporting InfiniBand and LVDS physical links.
ETA is designed for unattended operation, allowing configuration and recording to be controlled remotely.
\end{abstract}





\begin{document}

\begin{bottomstuff}
Author's address: C.~Patterson, 302 Whittemore Hall (0111),
Virginia Tech, Blacksburg VA
24061.\newline
E-mail: cameron.patterson@vt.edu
\end{bottomstuff}

\maketitle

\section{Introduction}
\label{sec:intro}

The Eight-meter-wavelength Transient Array (ETA) is a new radio telescope designed to observe a variety of postulated but as-yet undetected astrophysical phenomena which are suspected to produce single pulses detectable at relatively long wavelengths.
Potential sources for these pulses include the explosion of primordial black holes (PBHs) and prompt emission associated with gamma ray bursts (GRBs).
PBHs are postulated to arise from density fluctuations in the early universe rather than the gravitational collapse of a star.
If black holes evaporate as suggested by specific combinations of general relativity and quantum mechanics, those with masses below about $10^{14}$~g should be approaching a state of runaway evaporation, terminating in a massive burst of radiation \cite{hawking-nature}.
Also, prior to a possible terminal explosion, a primordial black hole may undergo a topological phase transition producing an observable outburst of energy if there exists a compact, extra spatial dimension in addition to the three known spatial dimensions [\citeNP{Kavic1a}b].

A number of models have been proposed to explain short-duration GRBs, including the merger of closely-separated compact objects such as neutron stars and/or black holes.
The formation of a single black hole would release an immense amount of energy over a few seconds \cite{grb}.
Prompt radio emission from GRBs would help to increase our understanding of these events.
Recently, a dispersed pulse of duration $<$ 5 ms was detected during the analysis of archival pulsar survey data \cite{Lorimer}.
The brightness (30~Jy, where 1~Jy = $10^{-26}$~W~m$^{-2}$~Hz$^{-1}$) and singularity of the burst suggest the source was not a rotating neutron star.
Although pulses can be quite strong by astronomical standards, they are difficult to detect due to their transient and unpredictable nature, and the narrow field of view provided by existing telescopes.

ETA, in contrast, is designed to provide roughly uniform (albeit very modest) sensitivity over most of the visible sky, all the time.
The complete array consists of 12 dual-polarized dipole-like elements (i.e., 24 radio frequency inputs) at the Pisgah Astronomical Research Institute (PARI), located in a rural mountainous region of Western North
Carolina ($35^{\circ}$ $11.98' $ N, $82^{\circ}$ $52.28'$ W).
Each dipole is individually instrumented and digitized.
The digital signals are combined to form fixed ``patrol beams'' which cover the sky, and the output of each beam is searched for the unique time-frequency signature expected from short pulses which have been dispersed by the ionized interstellar medium.
ETA has the computational resources and communication bandwidth to implement up to eight beamforming datapaths across all antennas, permitting it to operate as eight independent telescopes simultaneously recording data from different regions of the sky.
A patrol beam can be quickly reoriented by updating its beamforming coefficients.
Additional information about ETA and its science objectives are available at \cite{eta-website}.

This paper examines the science, engineering and computing aspects of the ETA project, and is organized as follows:
Section~\ref{sec:science} discusses some astrophysical objects of interest.
The ETA architecture, with particular emphasis on the digital back-end, is described in Section~\ref{sec:design}.
Section~\ref{sec:validation} discusses system validation issues such as determining bit error rates and ensuring channel synchronization.
The offline steps performed to mitigate radio frequency interference (RFI) and dedisperse pulses are given in Section~\ref{sec:analysis}.
Current status and conclusions are summarized in Sections~\ref{sec:status} and \ref{sec:conclusions}.

\section{Science Objectives}
\label{sec:science}

Traditionally, astronomical observations are directed at a specific astrophysical object.
With this goal in mind, astronomers use telescopes designed to selectively observe very small fields of view relative to the total sky visible above the horizon.
Often the objects of interest vary on astronomically long time-scales, so short temporal resolution is ignored in favor of the high signal-to-noise ratios obtained through long integration times.
There have been exceptions to this general approach.
For example, in the 1960s an array of dipole radio antennas was used to search for scintillating (``twinkling'') compact radio sources, leading to the discovery of pulsars \cite{Hewish}---neutron stars with rapid rotation periods of about 1~second, or less, observed by the periodic sweeping of their ``light-house'' beacons across the Earth.

Only relatively recently have astronomers begun systematic observations of the full-sky for short time-scale bursts of energy, or ``transients.''
These events may be produced by a host of highly energetic phenomena and are by nature unpredictable.
The progenitor object may be located anywhere and produce the event at any time.
To catch these events, observing systems are needed which run for extended uninterrupted intervals, and collect data from large factions of the total sky at any moment.
Satellite systems have discovered GRBs \cite{Klebesadel,Zhang}.
Optical telescopes have detected an  ``afterglow'' of GRBs \cite{Galama}.
Sources of potential, but as of yet unobserved radio transients include known phenomena (e.g., supernovae and GRBs), events expected to occur but not yet observed (e.g., mergers of two neutron stars), and exotic events (e.g., PBH explosions).

Below we discuss two phenomena that may produce an observable radio transient.
A GRB is an example of a known astrophysical phenomenon; observations of radio transients from GRBs would help constrain models of these events.
As an example of a more exotic phenomenon that could produce a radio transient we discuss the evaporation of a PBH in presence of an extra spatial dimension.
Observations of such an event would be of profound importance to our understanding of fundamental physics.

\subsection{Gamma Ray Bursts}

GRBs are the most luminous electromagnetic events in the observable universe.
They are pulses of gamma rays emanating from deep space which last anywhere from a few milliseconds to several minutes.
The initial burst is usually followed by a longer lived afterglow emitted at longer wavelengths (X-ray, ultraviolet, optical, infrared, and radio).
Most observed GRBs are thought to be a collimated emission caused by the collapse of the core of a rapidly rotating, high-mass star into a black hole.
While the radio afterglow has been observed, an unambiguous detection of prompt radio emission from a GRB has not yet been achieved.
Such detections would be useful in constraining and possibly identifying the nature of the progenitor object, the physics of the burst, and possibly the nature of the medium in which the burst took place.
A tantalizing possibility identified by \cite{Inoue} is that the prompt emission from GRBs at a redshift of $z \sim 10$ (and thus from the distant past) could be used as a probe of the Epoch of Reionization (the time during which objects, energetic enough to ionize neutral hydrogen, started to form in the early universe) and as a tool in constraining the nature of the intergalactic medium (IGM).  

There are a number of mechanisms which might result in a detectable radio component of the prompt emission from a GRB.
Benz and Paesold \citeyear{Benz} searched, without success, for radio transients produced within our Galaxy, and possibly observed in archived data obtained from radio telescopes observing solar radio emission.
Their interpretation of the data was based on a model for radio emission from an exploding PBH, as originally proposed by Rees \citeyear{Rees} and expanded by Blandford \citeyear{Blandford}.

Usov and Katz \citeyear{Usov} proposed that a low-frequency radio pulse of duration 1--100~s might result from the time variability of the current sheath surrounding a strongly magnetized jet flowing into the ambient interstellar plasma.
This model is interesting because under certain assumptions flux densities on the order of 10$^2$--10$^6$~Jy are possible at 30~MHz.
Successful observations would tell us about the radiating region in which the burst occurred and help to determine the mean intergalactic plasma density, through the dispersion measure (DM) for the pulse. 

While these scenarios offer limited prospects for detection using instruments with limited collecting area, work by Sagiv and Waxman \citeyear{Sagiv} renews hope that such detections may be possible with a small collecting area.
They propose that a strong maser-like coherent emission can emerge at the onset of shock deceleration with duration of about 1 minute.
The emission would be relatively narrowband with center frequency depending on the burst duration and various other poorly constrained parameters.
Inoue \citeyear{Inoue} finds that this emission may be about 1~Jy for a burst at $z\sim 1$, and such a detection would provide motivation to look for radio pulses from GRBs at greater redshifts.
Inoue shows that for $z < 3$ the dispersion of the pulse may be dominated by the local interstellar medium if the burst occurs inside or behind dense molecular clouds, possibly offering clues about the GRB host environment.
For $z > 3$ the dispersion would be dominated by the ionized IGM, therefore providing useful constraints on the medium.  For $z > 10$ the dispersion would in fact trace out the reionization history of the universe along the line of sight to the GRB, so that a number of detections would provide a detailed picture of the Epoch of Reionization. 

\subsection{Primordial Black Hole Explosions}

More exotic phenomena that could produce a radio transient, are associated with the evaporation of a PBH.
Two distinctly different outbursts are possible: a terminal explosive event, or an outburst that could occur as a PBH undergoes a topological phase transition in the presence of an extra spatial dimension.
An observation of either event would be of enormous importance.

As discussed by Hawking, a black hole should evaporate over time \citeyear{Hawking}.
The temperature $T$ of the black hole is 
\begin{equation}
T = \frac{\hbar c^3}{8\pi G k} \frac{1}{M},
\label{tem}
\end{equation}
where $M$ is the mass, and the emitted power is
\begin{equation}
P \propto \frac{\alpha(T)}{M^2},
\label{pow}
\end{equation}
where $\alpha(T)$ is the number of particle modes available for emission.
As the black hole evaporates the temperature and power increase, as does $\alpha(T)$, leading to the possibility of an explosive outburst.
Black holes of sufficiently small mass produced in the Big Bang (therefore ``primordial'') would now be reaching this final explosive phase.

Rees \citeyear{Rees} and Blandford \citeyear{Blandford} discussed in detail the coherent radio pulse that could be produced during such an event as a shell of electron-positron pairs emitted by the explosion expands into the surrounding interstellar medium.
The emitted radio pulse is the result of the current induced on the surface of the expanding shell as it excludes the ambient interstellar magnetic field from its interior.
To produce a coherent radio pulse the Lorentz factor of the relativistically expanding shell (the ``fireball'') must fall within the range $\gamma = (1-(v/c)^2)^{-1/2}\sim 10^5$ to $10^7$.
The expected transient pulse from such an event would be nearly 100\% linearly polarized, given the small length scale of the expanding shell compared to the length scale for irregularities in the interstellar magnetic field. 

Extra spatial dimensions, beyond the three known dimensions, have long been a subject of theoretical discussion in physics \cite{GSW}.
Typically, an extra dimension is circular (``toroidal''), having a periodic coordinate, and is of a small (``compactified'') currently undetected size.
If there is such a compactified, toroidal, fourth spatial dimension, an evaporating black hole will make a topological phase transition as its size shrinks to less than the length scale of the extra dimension.
The change in topology of the black hole is easy to understand.  A black hole larger in size than the length scale of the circular extra dimension will ``wrap'' the extra dimension, like a donut wraps along the circumference of a circle.
As the black hole evaporates and shrinks in size, it should unwrap and become localized within the extra dimension.
This topological phase transition will be accompanied by the emission of a fraction (of order 1\%) of the mass-energy of the black hole, as discussed by Kol \citeyear{Kol1b}.  

The analysis of Rees and Blandford can be adapted to this black-hole topological phase-transition scenario [\citeNP{Kavic1a}a; \citeyearNP{Kavic2}b].
Searches for transients of observed frequency $\sim$ 1~GHz to $10^{15}$~Hz will be sensitive to pulses from fireballs with Lorentz factors $\gamma \sim 10^5$ to $10^7$, and to possible extra dimension length scales $L \sim 10^{-18}$ to $10^{-20}$~m.
With an estimate of the distance to the evaporating black hole (available from the observed DM for an observed radio pulse), one can distinguish between a topological phase transition event and a final explosive event in which all the mass of the black hole is emitted (the event originally discussed by Rees and Blandford).

These extra dimension length scales have associated energy scales of $(L/\hbar c)^{-1}$ $\sim$ 0.1 to 10~TeV.
The Large Hadron Collider (LHC) will be sensitive to physics associated with a particular class of compactified extra dimension models up to energy scales of $\sim 1.5$~TeV \cite{PDG}.
Thus constructive comparison of astrophysical observations and LHC results would be possible.

In summary, a positive pulse detection of the sort produced by a topological phase-transition scenario could signal the presence of an extra spatial dimension, while a null detection would serve to constrain the possible size of an extra dimension.
The detection of a final explosive event in the evaporation of a PBH would also be of great importance, confirming a range of theoretical concepts at the boundary of gravitational and quantum physics.

\section{Instrument Design}
\label{sec:design}

ETA is designed to operate in the range 29--47~MHz, which is a response to a number of factors.
First, some astrophysical theories suggest the possibility of strong emission by the sources of interest in the HF and lower VHF bands, limited at the low end by the increasing opacity of the ionosphere to wavelengths longer than about 20~m (15~MHz).
Useable spectrum is further limited by the presence of strong interfering man-made signals below about 30~MHz (e.g., international shortwave broadcasting and Citizen's Band (CB) radio) and above about 50~MHz (e.g., broadcast television), which make it difficult to observe productively outside this range.
At these frequencies, however, the ubiquitous Galactic synchrotron emission is extraordinarily strong and can be the dominant source of noise in the observation \cite{antennas}.

\begin{figure}[t]
    \begin{minipage}[b]{1.0\linewidth}
        \centering
        \centerline{\epsfig{figure=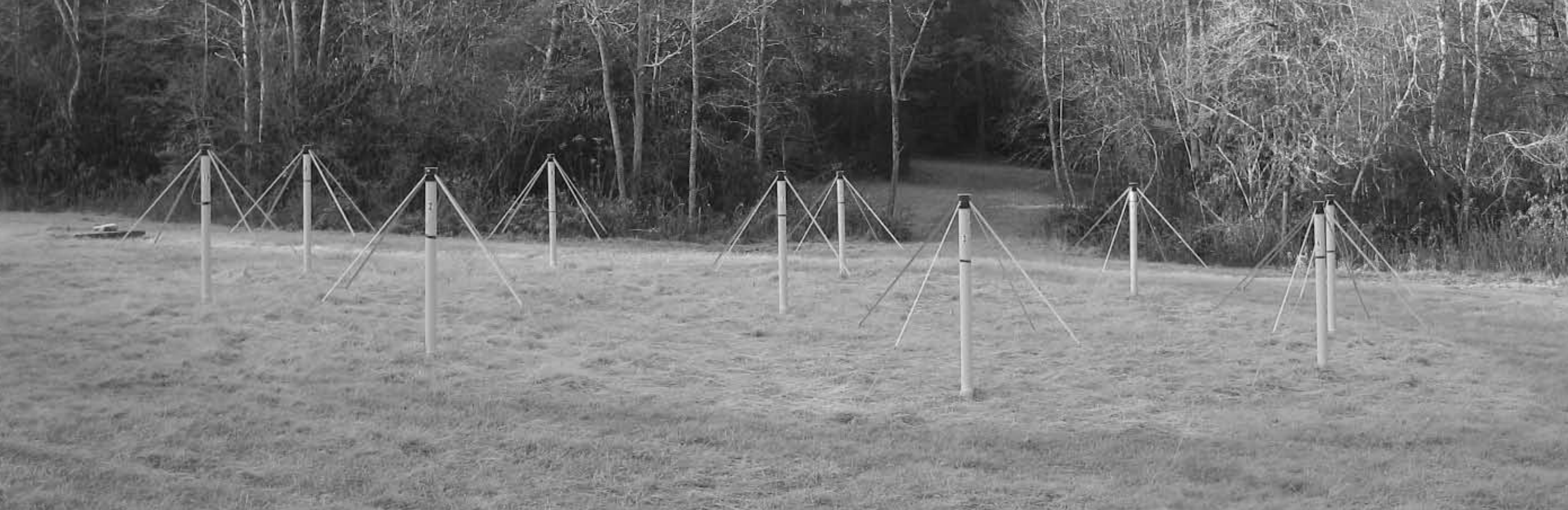,width=\linewidth}}
    \end{minipage}
    \caption{ETA's ten-element core array.  One antenna stand is in the center, and the remaining nine stands form a circle of radius 8~m.  At each 2~m high stand, two dipoles are connected to an active balun (dipole-to-coaxial converter and preamplifier) located inside the PVC mast.}
    \label{fig:antenna_array}
 \end{figure}

\begin{figure}[t]
    \begin{minipage}[b]{1.0\linewidth}
        \centering
        \centerline{\epsfig{figure=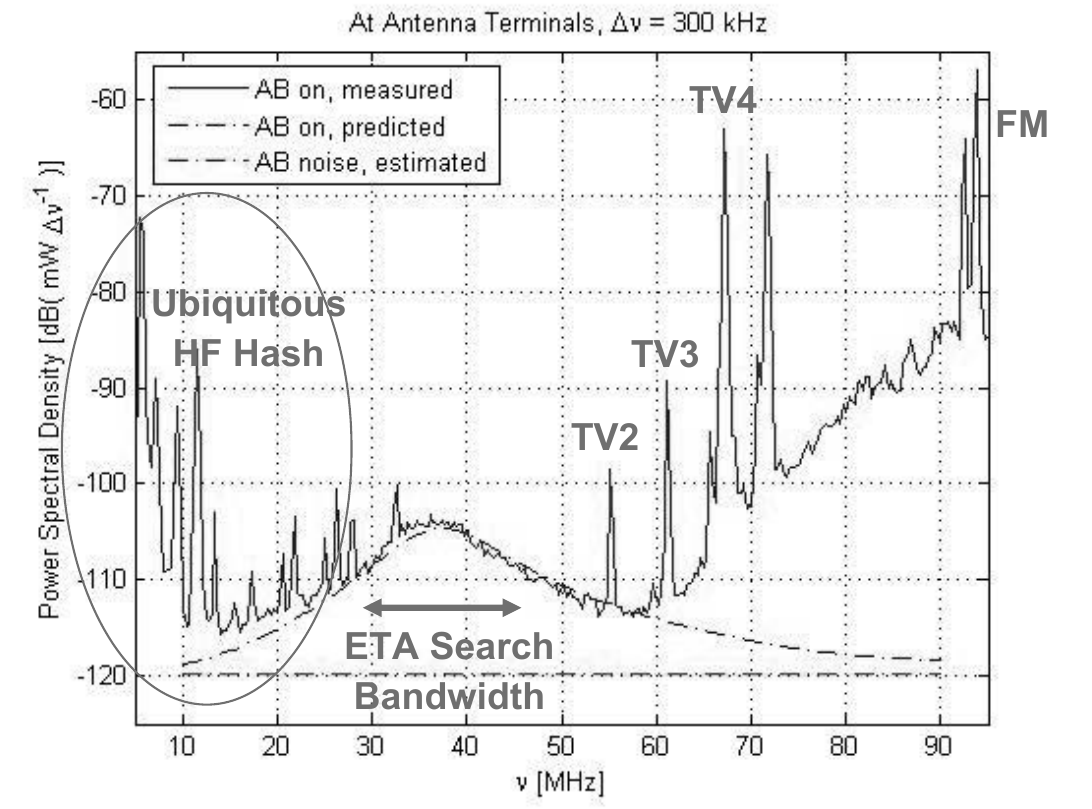,width=84mm}}
    \end{minipage}
    \caption{A typical power spectrum measurement for an ETA dipole's active balun (AB).  Over the 29--47~MHz range, sensitivity is limited only by Galactic noise (the upper dashed line) and not AB noise (the lower dashed line).}
    \label{fig:spectrum}
 \end{figure}

\begin{figure}[t]
    \begin{minipage}[b]{1.0\linewidth}
        \centering
        \centerline{\epsfig{figure=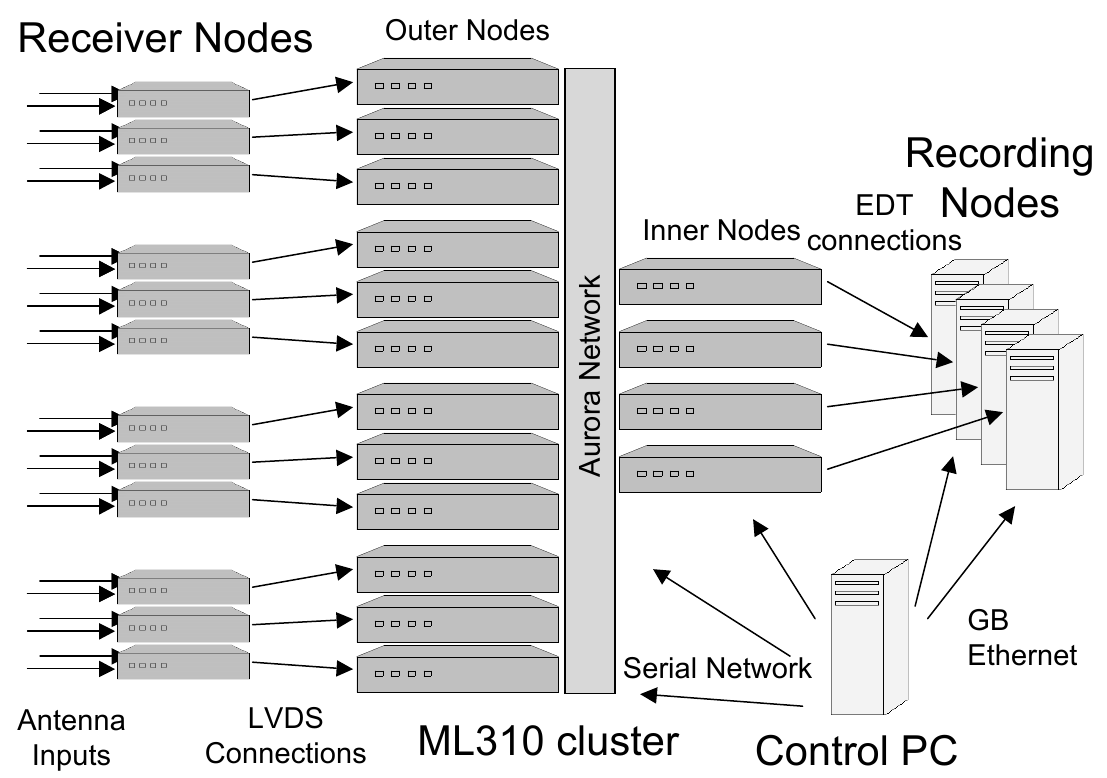,width=100mm}}
    \end{minipage}
    \caption{ETA's digital front-end and back-end.}
    \label{fig:system_arch}
 \end{figure}

The ETA system receives RF input from 24 dipole-like antennas mounted on 12 stands.
Figure~\ref{fig:antenna_array} shows the ten-stand core array; not shown are a pair of outrigger stands located several hundred feet away to the north and east.
Each stand supports two orthogonally-polarized, dipole-like elements.
Figure~\ref{fig:spectrum} is a snapshot of typical spectrum seen by the antenna.
Buried coaxial cable connects antenna outputs to analog receivers that amplify and filter the signals.
The digital signal processing and data recording are the focus of this paper, and are shown in Figure~\ref{fig:system_arch}.

The architecture consists of three tiers: receiver nodes, FPGA cluster nodes, and data acquisition nodes.
At each level, data must be synchronized and processed in parallel.
The system receives antenna signals through the receiver nodes consisting of twelve Altera Stratix DSP development boards, referred to as S25s \cite{dsp-s25}.
Each S25 board performs A/D conversion and digital filtering for two dipole inputs, and provides a source-synchronous stream of data to a cluster node.
The cluster nodes are Xilinx ML310 FPGA boards, referred to as ML310s \cite{ml310}, and form twelve outer and four inner nodes networked together.
Cluster nodes process, combine, and prepare data for recording on the four Dell SC430 server-class PCs.
Stored data may be analyzed while the system is not acquiring data, or archived to tape.
A fifth PC controls and monitors system functions.
Sections~\ref{ssec:receiver} and \ref{ssec:fpgacluster} describe the instrument's front-end and back-end in greater detail.

\subsection{Receiver Nodes}
\label{ssec:receiver}

The main functions of the receiver nodes are to digitize, downconvert and channelize (filter to a narrower bandwidth) the analog antenna input.
Since designing custom hardware to perform this function would be costly and time consuming, this functionality was implemented on commercially available Altera S25 boards.
This board was chosen because it contains the necessary hardware elements for the design, and was familiar to the design team.
Each S25 board contains two 12-bit 125~MSPS A/Ds, allowing all 24 analog input signals to be digitized with 12 boards.
Approximately 80\% of the EP1S25 FPGA's 25K logic elements are used.

To maintain synchronization across all receiver nodes, one board generates the clock and reset signals for the other S25 boards.
A counter, synchronous across all S25 boards, is encoded with each time sample.
This counter is used in the cluster nodes to align data streams from different S25s.
Two antenna streams are combined into a single stream and transmitted as LVDS signals to a cluster outer node at 30 MB/s.
The combined stream consists of a 60~MHz clock, four bits of source synchronous data, and a counter bit.

\subsection{FPGA Cluster}
\label{ssec:fpgacluster}

\begin{figure}[t]
   \begin{minipage}[b]{0.55\linewidth}
        \centering
        \centerline{\epsfig{figure=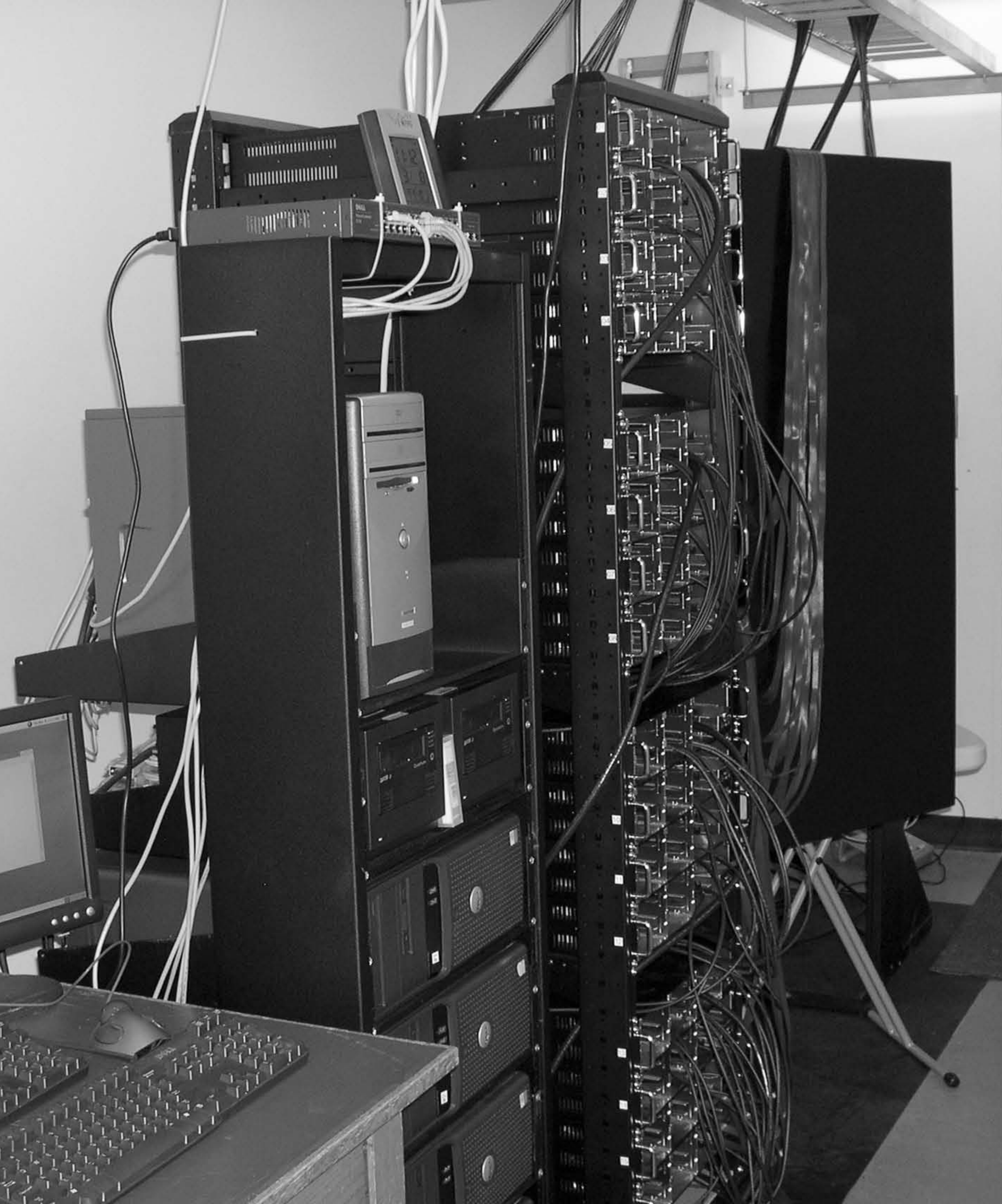,width=68mm}}
    \end{minipage}
    \begin{minipage}[b]{0.45\linewidth}
        \centering
        \centerline{\epsfig{figure=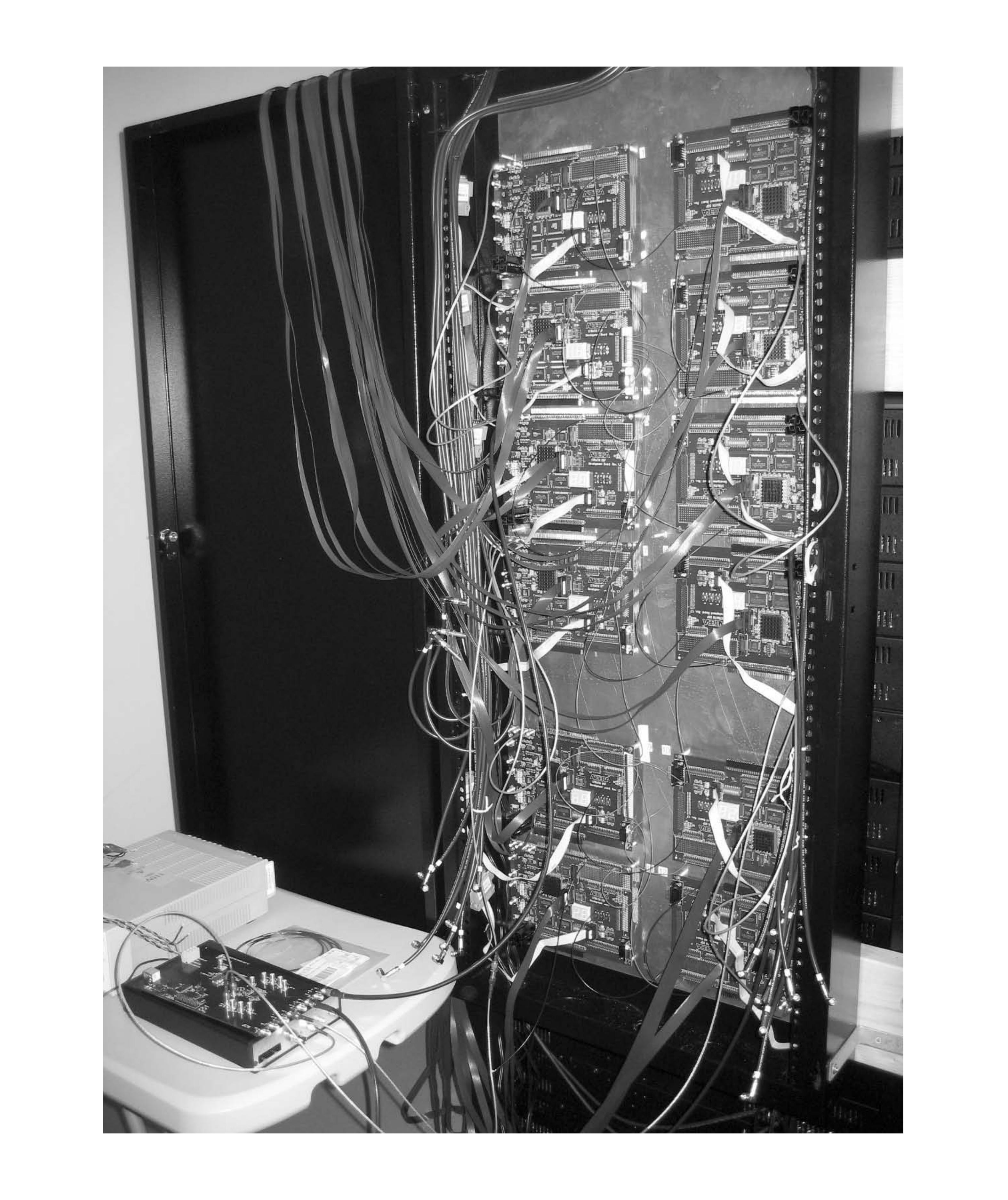,width=68mm}}
    \end{minipage}
    \caption{Inside ETA's equipment hut.  The larger picture shows the 16-node ML310 cluster occupying the middle rack between the PC cluster and tape drives on the left and the receiver node cabinet on the right.  The smaller picture shows the twelve S25 boards mounted inside the cabinet.}
    \label{fig:rack}
 \end{figure}

The FPGA computing cluster consists of 16 ML310 boards interconnected with 1X InfiniBand cable assemblies supporting a 2.5~GHz signaling speed.
Like the S25 board, the ML310 was chosen because of it was familiar to the design team and provides sufficient hardware resources and connectivity.
Cluster nodes are divided between 12 outer nodes, which connect directly to the receiver nodes, and 4 inner nodes, which connect directly to the PCs.
The FPGA cluster provides a versatile, synchronous, high speed network for merging antenna streams to support operations such as FFTs,  beamforming and RFI mitigation, and reformatting either raw or reduced data for transfer to the PC recording nodes.
In fact the FPGA cluster is as much a router as a computing engine.
The cluster rack is shown in Figure~\ref{fig:rack}.

\begin{figure}[t]
    \begin{minipage}[b]{1.0\linewidth}
        \centering
        \centerline{\epsfig{figure=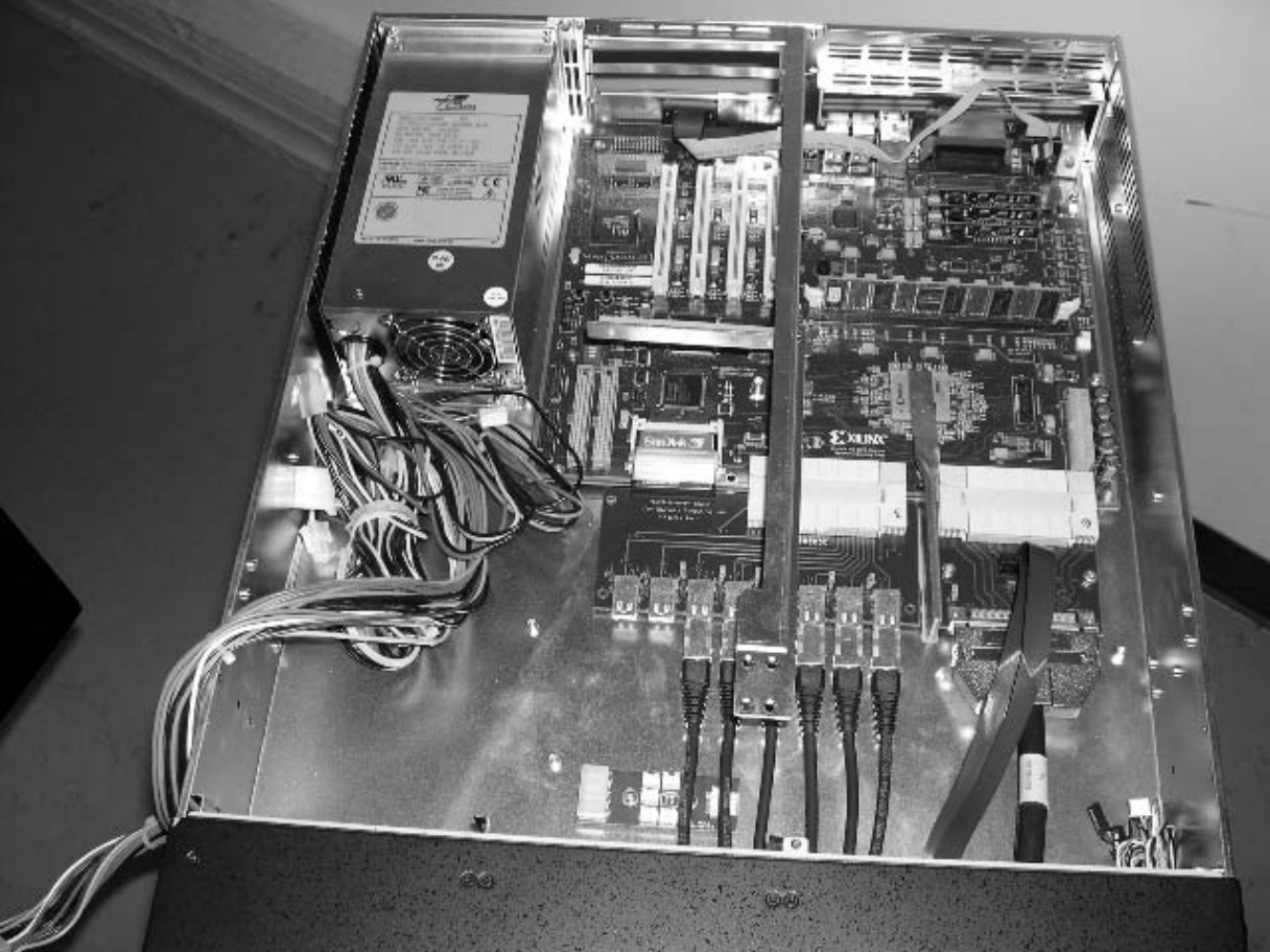,width=\linewidth}}
    \end{minipage}
    \caption{ML310 board and adapters.  The ML310's form factor allows mounting inside PC ATX motherboard cases.  This inner node has six InfiniBand cables connected to the adapter board on the bottom left of the ML310.  The adapter board to the right provides a MICTOR connector to a 3~m Precision Interconnect ``Blue Ribbon'' coaxial cable coming from an S25 receiver node, and an Amphenol 80-pin right angle connector to a 7~ft black cable going to an EDT PCI CDa data acquisition card in a PC node.  Between the two adapter boards is a copper bar conducting heat from the FPGA to the bottom of the case.}
    \label{fig:ml310}
 \end{figure}

The ML310 board shown in Figure~\ref{fig:ml310} has a Xilinx 2VP30 Virtex-II Pro FPGA, 256~MB DDR DIMM, 512~MB CompactFlash card, standard PC ports, and personality modules interfaces for RocketIO and LVDS access.
Custom adapter boards were designed for the personality module interfaces.
An adapter board provides InfiniBand HSSDC2 connector access to the FPGA's eight RocketIO transceivers.
These connections implement the lightweight Aurora link-layer, point-to-point protocol \cite{aurora}.
A six-layer adapter board provides a MICTOR connector interface to an S25 receiver node, and a source synchronous, 16-bit parallel LVDS cable interface to a PC node's EDT PCI CDa LVDS/600E data acquisition card \cite{edt}.
Each ML310 also communicates with the control PC through a UART, allowing all cluster nodes to be configured and queried remotely.

Routing in this architecture is simplified since the data flows only from the outer to the inner nodes.
Outer nodes are responsible for a large portion of the beamforming computation, while the inner nodes synchronize and combine the signals produced from the outer nodes.
Each inner node receives an input stream from up to six outer nodes, combines them, and outputs the result to a PC.
The upper and lower halves of the FPGA and PC cluster are independent and process antenna input from the north-south and east-west polarizations respectively.

ETA uses all three forms of clocking methodologies: system-synchronous within receiver, inner and outer nodes; source-synchronous for data transmission between S25s and ML310s, and between ML310s and the PC's data acquisition cards; and self-synchronous for transfers between outer and inner nodes \cite{serialiobook}.
The ML310's receiver node interface uses the 60~MHz clock sourced by the S25 to synchronize data transfers.
All other inner and outer node modules use the ML310 board's 125~MHz reference clock, which is divided by 2 for the beamforming datapath, and multiplied by 20 in the RocketIO transceivers to obtain a bit-rate clock for the serializer and deserializer.
The transmitted signal encodes the data such that the clock can be recovered by the receiver.
Brief resynchronizations referred to as clock corrections occur periodically, necessitating buffers on the transmitters and receivers.

\begin{figure}[t]
    \begin{minipage}[b]{0.5\linewidth}
        \centering
        \centerline{\epsfig{figure=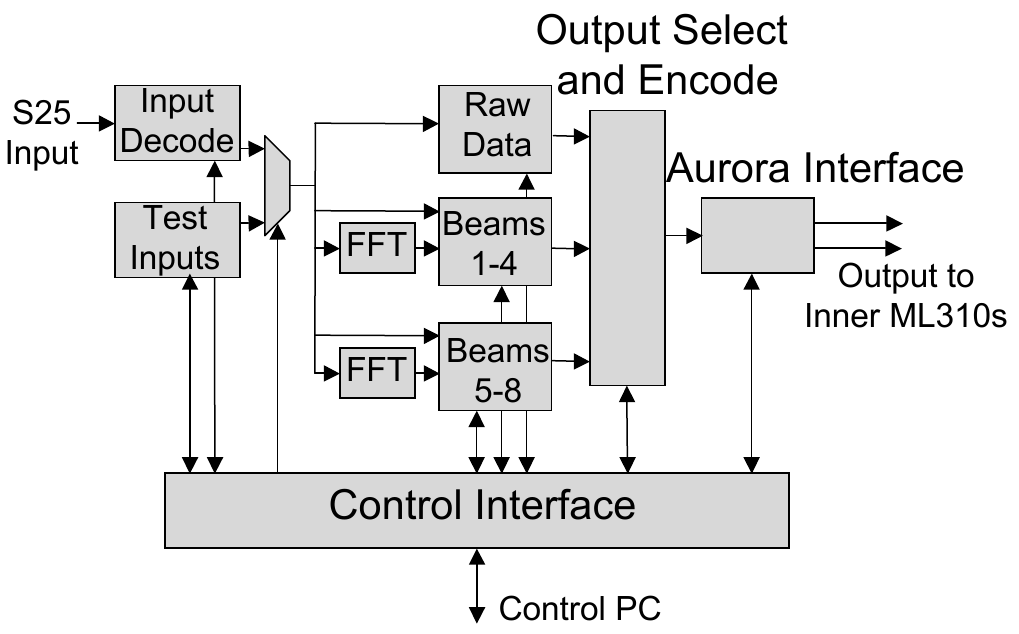,width=60mm}}
        \center{(a) Outer node.}
    \end{minipage}
    \begin{minipage}[b]{0.5\linewidth}
        \centering
        \centerline{\epsfig{figure=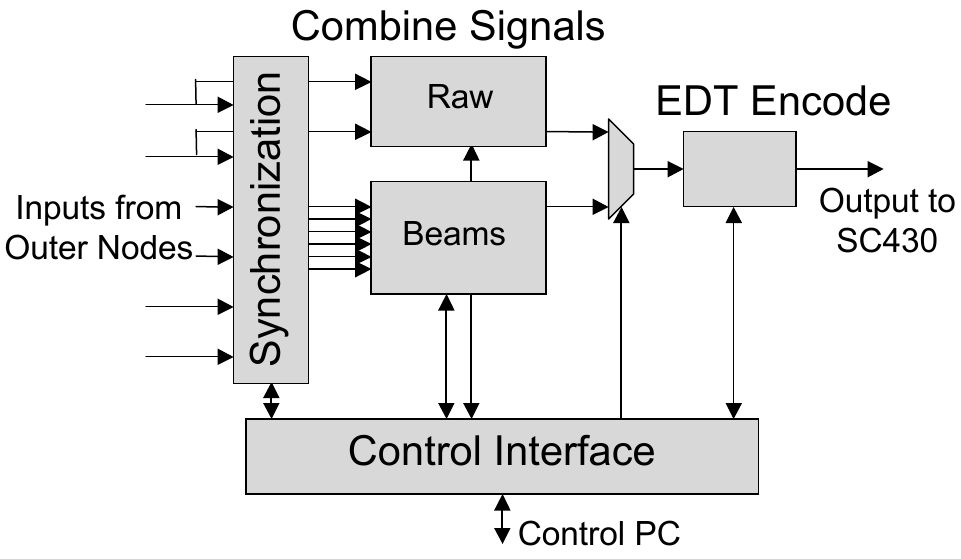,width=60mm}}
        \center{(b) Inner node.}
    \end{minipage}
    \caption{ML310 implementations.}
    \label{fig:ml310-nodes}
 \end{figure}

Raw data collection mode allows the output from eight of the S25 boards to pass directly to the PCs.
Each outer node receives data from two antennas at a combined rate of 30~MB/s, reformats the data, and passes it to an inner node using the Raw Data path shown in Figure~\ref{fig:ml310-nodes}(a).
The inner nodes illustrated in Figure~\ref{fig:ml310-nodes}(b) collect data from two outer nodes, combine and format the input streams for storage on the PC cluster.
In this mode each of the four PCs receives and store data at 60~MB/s, giving the system an aggregate recording rate of 240~MB/s.
Data collections usually occur for roughly one hour, generating an 800~GB dataset.
After offloading the data to 400~GB LTO-3 tapes, another acquisition can begin.

The basic beamforming mode uses all 12 outer nodes, combining multiple streams of data to form patrol beams for improved sensitivity.
Although the sky could be tessellated with up to 12 independent beams, only 8 beams are implemented due to hardware resource limits and poor sensitivity (because of antenna pattern roll-off) near the horizon.
Each outer node again receives data from two antennas at a rate of 30~MB/s, or 360~MB/s aggregate.
The outer nodes multiply each antenna input stream with the corresponding beamforming coefficients and sum the results to produce eight single-pol beams.
Four of the beams are sent to one inner node and the other beams are sent to an adjacent inner node as illustrated in Figure~\ref{fig:ml310-nodes}(a).
At this stage each Aurora connection is transmitting 120~MB/s for an aggregate bandwidth of 2.88~GB/s to the inner nodes.
Each connection has an available bandwidth of 240~MB/s (5.76~GB/s aggregate), increasable to 360~MB/s (8.64~GB/s aggregate) by selecting a faster RocketIO reference clock although the bit error rate (BER) will increase.
Each inner node combines a four-beam input from six outer nodes.
The six input streams are synchronized, with corresponding samples summed, shifted and rounded (to avoid a DC bias) before transmission to the PCs.
This reduction allows each PC to record four single-pol beams at 60~MB/s, or an aggregate 240~MB/s.
For each polarization, one PC contains data for beams 1 to 4 and the other PC contains data for beams 5 to 8.

Since propagation speed through the ionized interstellar medium varies with the frequency of the wave, pulse dispersion or smearing results.
In order to dedisperse the signal, the observing band is split into narrow channels and the detected signal from each channel is delayed by a different amount before summation to obtain the total power signal.
Even on a fast workstation, dedispersion requires over twenty-four hours for each hour of data collected, with the FFT operations accounting for roughly half of this time.
Because it is difficult to completely automate pulse detection, the main focus is on speeding up dataset post-processing.
The FPGA cluster can optionally perform a real-time, 1024-point FFT of the 18~MHz-wide passband into 1024 17.6~KHz-wide spectral channels with 66~$\mu$s integration.
Beamforming is applied on a per-channel basis, leaving the software to apply a range of DMs.

\begin{figure}[t]
    \begin{minipage}[b]{1.0\linewidth}
        \centering
        \centerline{\epsfig{figure=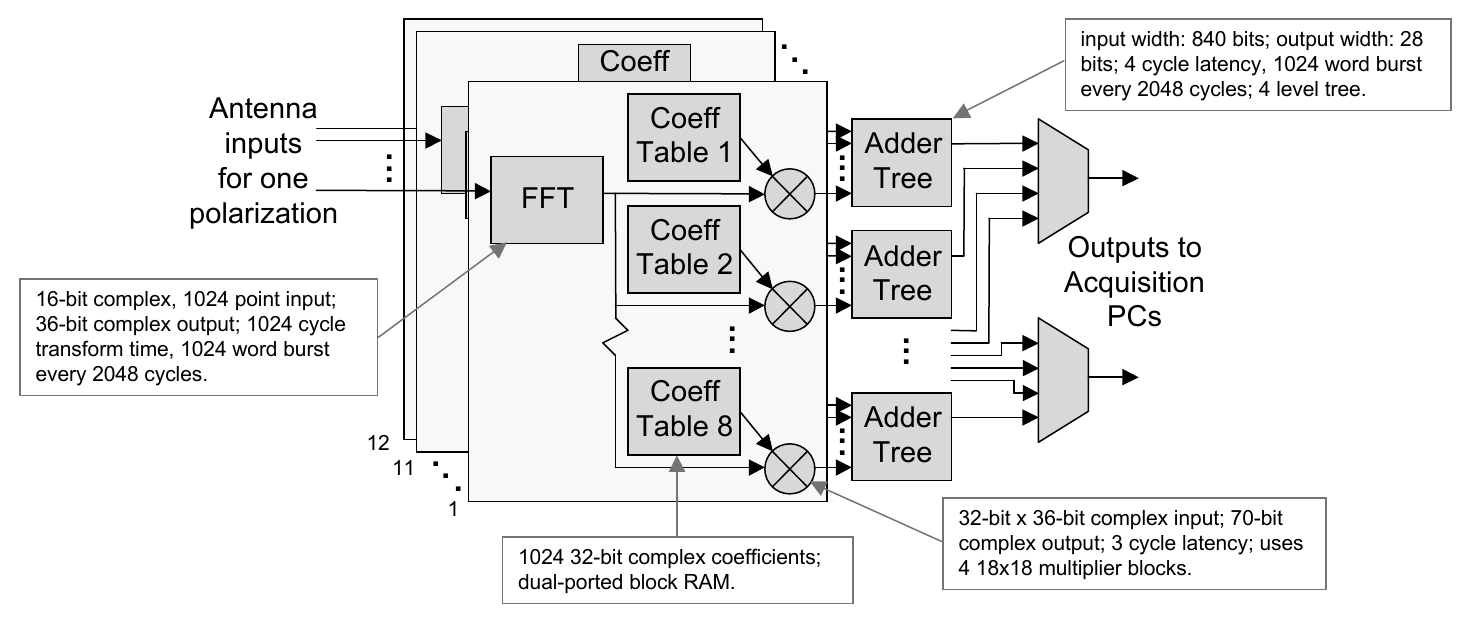,width=\linewidth}}
    \end{minipage}
    \caption{Beamforming across spectral channels.}
    \label{fig:spectral}
 \end{figure}

The outer nodes buffer 1024 samples which are streamed into the FFT block.
FFT computations requires 1024 clock cycles, after which the output is streamed to the beamforming stage as indicated in Figure~\ref{fig:spectral}.
Each beam must multiply the 1024 frequency domain values with corresponding complex coefficients.
For the two input streams and the eight beams formed, each outer node contains a table with 16384 coefficients loaded through the UART.
After the multiplication, corresponding beams are summed and multiplexed for transmission to the inner nodes as in the basic beamforming mode.
A useful feature of the FFT mode is that frequency bins can be selectively included or removed, enabling the system to record only the frequencies of interest and extend viewing time.
 
As shown in Figure~\ref{fig:ml310-nodes}(b), the main functions of the inner node are to synchronize the streams from the outer nodes, combine them, and format the output for storage on a PC.
The streams are synchronized with a bit field indicating the start of a vector.
Corresponding entries from a set of vectors are summed and shifted to extract the appropriate bits.
A 32-bit value encodes the 4-bit vector start flag and either one or two complex 14-bit samples, depending on the mode.
Two 32-bit streams are multiplexed and output to a PC.
This allows a sustained 60~MB/s transfer to disk of either four antenna streams or beams, and a system maximum of sixteen antenna streams or single-pol beams.
Sixteen (i.e.\ eight dual-pol) beams allows a majority of the viewable sky to be observed with maximum sensitivity.
  
The outer nodes utilize approximately 65\% of the 30K logic elements, 44\% of the 136 18-Kbit BRAMs, and 65\% of the 136 18$\times$18 multipliers in the 2VP30 FPGA.
By contrast, less than 28 (5\%) of the  IOBs are required.
Both inner and outer nodes use six of the eight available gigabit transceivers, necessitating a lightweight protocol such as Aurora.
Excluding the FFT, each outer node performs 2~GMAC/s while beamforming is active, for a system aggregate of 24~GMAC/s.

\section{System Validation}
\label{sec:validation}

To enable BER testing, synthetic data can be sourced either from the S25 or ML310 nodes.
The data (typically counter or numerically-controlled oscillator output) are verified before and after every communication link in the system, and any errors are tallied on each ML310.
The serial connection between each ML310 and the control PC allows for internal state to be read or written.
C programs convert user-generated text files to a bit string defining the operating mode and data such as beamforming coefficients, or from a bit string to text files describing the internal state.
Readback also returns the status of the Aurora links, indicates whether buffers have ever overflowed, and confirms the beamforming coefficients were written correctly.
Error counts may be queried after a test to facilitate on-site or remote error rate testing and diagnosis, or after a data acquisition to check that the integrity of the data collected.
In addition, synthetic data tests may be invoked at the end of a data collection script.
Through the detailed information received from the ML310s, problems can be quickly isolated to specific connections, cables or boards.
These scan paths allowed each half of the FPGA cluster to be connected and tested in less than a day.

ETA's systolic architecture makes data retransmission difficult, and error correcting codes would be a significant overhead given the number of communication links.
However, digital system errors are unacceptable even though RFI is a much larger concern.
All data errors originating in the ETA system have occurred in the 2.5~Gbps Aurora channels.
Synchronization errors (manifesting as detectable buffer overflows) have never been observed.
Careful design and signal integrity analysis of the adapter boards, and transmitter pre-emphasis and differential voltage swing adjustments, have resulted in observed BER less than the InfiniBand maximum of $10^{-12}$.
In a 200~GB dataset, the number of data bit errors is generally 0, with a maximum of 2 observed.
The main source of errors are the connections between a cable and an adapter board.
Outer nodes transmit only two channels and make each channel available on three of the adapter board connectors to facilitate bypassing a connection generating errors.
Inner nodes use all six input channels, and errors are usually remedied by reseating cable connections.
Enabling the RocketIO transceiver's CRC feature and checking if any CRC errors have occurred during an observation should address inter-board data integrity concerns.

\section{Offline Data Analysis}
\label{sec:analysis}

A typical ETA observation is about 1 hour long.
In beamforming modes, the output of each beam is captured as 200 files 1~GB in size, where each file consists of about 17 seconds of 14 bit (7 bits ``I'' $\times$ 7 bits ``Q'') data sampled at 7.5~MSPS, plus additional bits inserted by the real-time digital hardware for the purpose of diagnostics and synchronization.
So, for example, a 4-beam observation results in 800~GB of data.  In per-antenna modes, the format is exactly the same, except each set of 200 files represents 1 antenna.  

We now describe how this data is analyzed to detect dispersed pulses.
The first step is a data integrity check.
The intent of this step is to identify data which is corrupted and unusable either due to internal system errors, or due to excessive levels of RFI.
Diagnostic counter bits are checked to confirm continuity (i.e., no skipped, missed, or garbled counters) within each file and between files.
As explained above, such errors are exceedingly rare and we have never detected a counter error indicating loss of synchronization.
The statistics of the data are then analyzed to confirm that the number of ``clips'' (extreme data values) is reasonable and that the magnitude statistics of the ``I'' and ``Q'' samples are both approximately Gaussian with the same variance and with acceptably low DC offsets.
This procedure is useful for identifying times when the data stream is RFI-dominated, so that this can be taken into account when selecting time intervals for performing baseline calibration and other RFI-sensitive computations, discussed below.

\subsection{Spectrogram Generation and RFI Mitigation}

The next step is to create ``raw'' spectrograms.
The data is partitioned into 1K sample blocks.  Each block is individually transformed into the frequency domain using a 1K FFT.
The FFT is preceded by a Bartlett (triangular) window, which serves to mitigate edge effects---a particularly acute problem when RFI is present---while not giving up too much frequency resolution.
The resulting spectrograms have time-frequency resolution of 136.5~$\mu$s and 7.324~kHz, respectively.
Sensitivity is improved by incoherently averaging FFT outputs such that the time-frequency resolution is matched to either the DM of the source under study; or, for DM searches, the DM being tested.
For example, this is achieved for the Crab Pulsar (DM $= 56.8~\mbox{pc}/\mbox{cm}^3$) by averaging FFTs to a time resolution of 8.738~ms.
This averaging also serves to mitigate impulsive RFI, which is often resolved at 136.5~$\mu$s but which has low very duty cycle and so is significantly suppressed by this kind of averaging.

At this point, it is desirable to remove the frequency response of the instrument (in particular, the bandpass shape) and the time-domain variation in total noise power which is due to the natural diurnal variation of the Galactic background.
We select one spectrogram every 7.5~minutes to use as a reference in this calibration.
If a selected spectrogram exhibits excessive RFI (as noted in the first step), then adjacent spectrograms are examined until a sufficiently-RFI-free set of spectrograms is found.
Each spectrogram is averaged over 17~s (the length of the orginal file) to obtain a frequency response (including the time-variable magnitude) at that time.
We then interpolate over this sparse set of frequency responses to obtain the desired ``baseline spectrograms.''
Calibration consists of dividing the raw spectrograms by the baseline spectrograms, nominally yielding new spectrograms with flat frequency response with mean value 1 in both time and frequency.

RFI mitigation operates directly on the calibrated spectrograms.
First, the top 1\% of all pixels are set equal to one (i.e., the mean pixel value); this alone is effective in mitigating much of the RFI, and removes outliers that would otherwise bias statistical measures required in subsequent steps.
Next, we compute the standard deviation of pixel magnitudes ($\sigma_{tf}$), the standard deviation of the total-power time series ($\sigma_{t}$), and the standard deviation of the integrated spectrum ($\sigma_{f}$).
Then, we set equal to one: (1) any pixel which is greater than $3\sigma_{tf}$, (2) any pixel occuring at a time associated with a total-power time series value greater than $3\sigma_{t}$, and (3) any pixel occuring at a frequency associated with a integrated spectrum value greater than $3\sigma_{f}$.
The number of additional pixels modified as a result of this procedure is typically not much greater than the 1\% modified in the first step. 

\begin{figure}[t]
    \begin{minipage}[b]{1.0\linewidth}
        \centering
        \centerline{\epsfig{figure=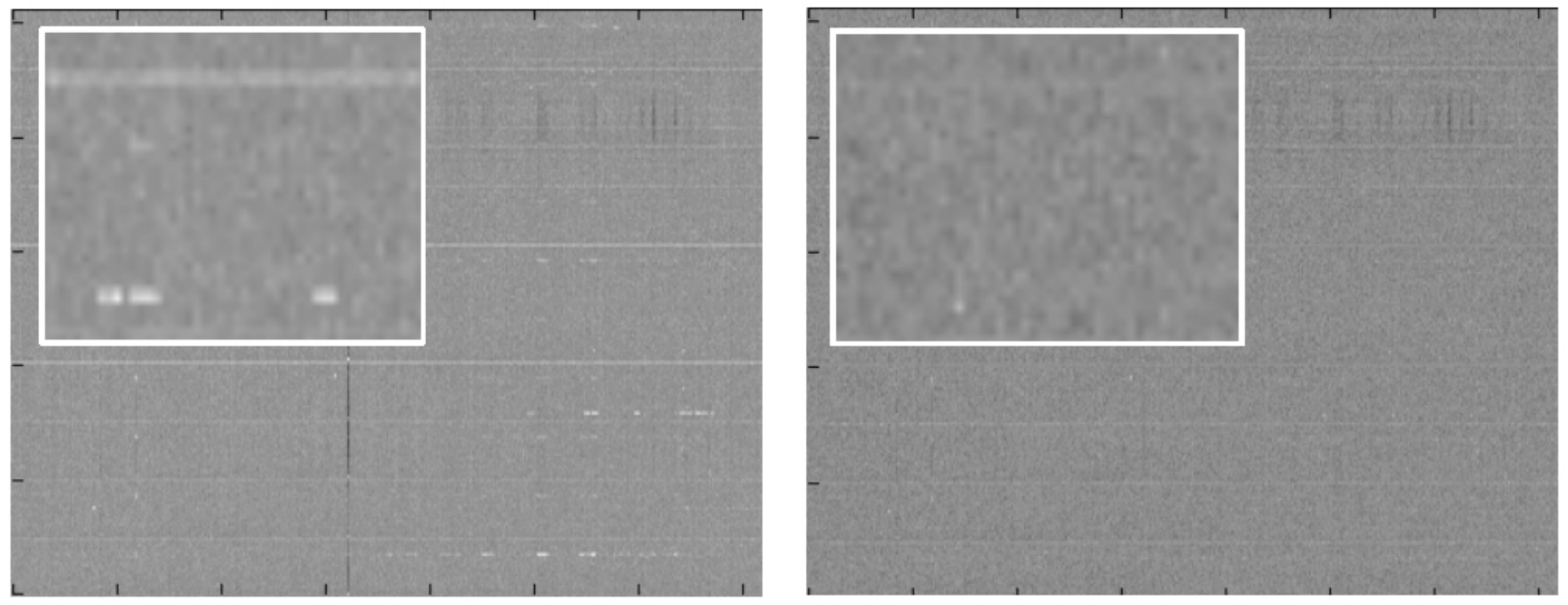,width=\linewidth}}
    \end{minipage}
    \caption{A typical observation in spectrogram form, prior to dedispersion.
    The left and right images are before and after RFI mitigation, respectively.
    Inset views zoom in to $1/8$ the time/frequency span of the larger images.
    Vertical dimension is frequency: 38 $\pm$ 1.875~MHz with 7.324~kHz resolution. 
    Horizontal dimension is time: 1 hour with 498~ms resolution (integrated from 8.738~ms to reduce image size).}
    \label{fig:spectrograms}
 \end{figure}

Figure~\ref{fig:spectrograms} shows typical data before and after RFI mitigation.
The left plot shows several forms of RFI that we commonly contend with, including persistent narrowband RFI, short-term broadband RFI, and various ``hybrid'' forms including the band of ``scalloping'' along the top edge and narrowband bursts appearing in the bottom right.
This RFI would ordinarily saturate the subsequent dispersed pulse search.
Note the RFI is mostly absent in the right plot, although the scalloping along the top edge and a few other anamolies remain.
This level of RFI is sufficiently low to facilitate dedispersion and dispersed pulse searching. 

\subsection{Dedispersion and Pulse Detection}

Dedispersion proceeds as follows:  The DM of interest determines the extent to which high frequencies must be delayed relative to low frequencies through a simple formula.
This is implemented in the calibrated spectrograms simply by shifting the low-frequency ``rows'' the appropriate number of pixels with respect to the high-frequency rows.
The dedispersed time-domain total-power signal is then obtained simply by averaging over frequency bins.
This method is known as ``incoherent dedispersion,'' because phase information is not used or preserved and thus the resulting sensitivity is somewhat less than the optimal method, which is known as ``coherent dedispersion.''
The prevelence of RFI combined with the availability of a relatively simple incoherent method of RFI mitigation make incoherent dedispersion desirable in our case.   

The resulting dedispersed time series is then downsampled to be well-matched to the expected scatter-broadening of the pulse.
This is to account for the refraction within the insterstellar medium, which ``smears'' the pulse to be many times longer than its duration at the source.
For example, a dedispersed Crab giant pulse is expected to be scatter-broadened to a duration somewhere between 0.5~s and 5~s, depending on the interstellar ``weather,'' which itself shows significant time-variability.
Thererfore, when observing the Crab, we downsample from 8.738~ms to 498~ms, which greatly improves sensitivity.  

Pulse detection is then performed as follows:  The mean and variance of the dedispersed time series are determined, and then the largest-valued samples determined.
In general, samples need to be at least $5\sigma$ greater than the mean to be considered significant.
Because there is large uncertainty in the scatter-broadened length of pulses, the procedure is repeated after decimation by averaging adjacent samples, yielding a search over powers-of-two in scatter-broadened duration.
The maximum duration considered is currently 8~s.
Detected pulses are then subjected to a follow-up procedure in which the data are reprocessed for a range of DM about the discovery DM.
This is to confirm that the quality of detection degrades with increasing DM offset, indicating that the pulse truly is dispersed as expected and is not simply some artifact of the processing.  

Our current best pulse detection is of a Crab giant pulse, which was detected using the above procedure at a level of about $4\sigma$.
Although the level of significance is below our desired standard of $5\sigma$, it was confirmed using the DM sweep procedure described in the previous paragraph that the detection peaks sharply at the expected DM.
Also, the calibrated flux density estimate of 900~Jy (38~MHz) for this detection is quite consistent with recent reported detections (by much larger telescopes) at 23~MHz and 200~MHz.

\section{Current Status}
\label{sec:status}

\begin{figure}[t]
    \begin{minipage}[b]{1.0\linewidth}
        \centering
        \centerline{\epsfig{figure=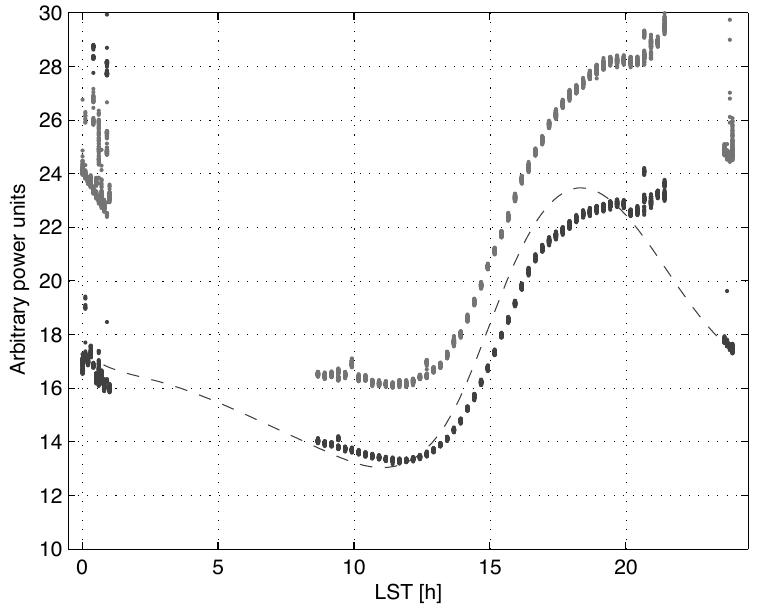,width=84mm}}
    \end{minipage}
    \caption{Diurnal variation in a 5~MHz bandwidth centered at 45~MHz.  The dotted line is predicted power due to galactic noise, while the black and gray scatters are measured power from a core array dipole and the east outrigger respectively.}
    \label{fig:diurnal}
 \end{figure}

ETA development began in August 2005, with a demonstration of direct sampling from the first four dipoles three months later.
One commissioning test was confirmation of the diurnal variation shown in Figure~\ref{fig:diurnal}.
The first 200~GB raw data acquisition through S25 and ML310 nodes to a PC occurred in April 2006.
A lightning strike in July 2006 damaged preamplifiers, but most have since been repaired.
FFT beamforming mode was added in July 2007.
By early 2008 over 16~TB of raw data has been collected, largely consisting of one-hour contiguous observations taken at a constant sidereal time for use in the PBH search.
Two independent dedispersion software tool chains have been developed, and we see a role for undergraduate students in applying these tools to the dataset archives.
An analysis of how frequently pulses are detected and their DMs should provide valuable insights to the distribution of the progenitor objects.

Mitigation of RFI is an important consideration for any instrument operating at low frequencies.
Although dedispersion tends to suppress narrowband RFI, and FFT bins may be selectively blanked to reject noise in certain frequency ranges, an additional approach to RFI rejection is anticoincidence.
A portable second instrument (ETA2) is being assembled at a site more than 300~km from PARI.
ETA2 still sees the same section of the sky as ETA but is not affected by ETA's local RFI.
Although each station records time accurately via GPS, there is no beamforming across stations since data collection is not synchronized during the observation.
During dataset post-processing, pulses detected simultaneously at both ETA and ETA2 would rule out local RFI sources.

\section{Conclusions}
\label{sec:conclusions}

FPGAs with integrated multi-gigabit serial transceivers are ideal platforms for the signal processing requirements of radio telescope arrays.
Synchronous streaming computations may be implemented over many boards without requiring system-wide clock distribution.
As in ETA, the physical communication topology can be tailored to the dataflow requirements, simplifying the interfaces and development effort.
These characteristics also make FPGAs containing transceivers an essential technology in particle accelerators such as the LHC \cite{LHC-ATLAS}.
In contrast, software processors and their communication protocols are difficult to use in the lower tiers of telescope and accelerator instruments due to variable latencies.

Our goal to obtain results within the first year necessitated the use of commercial-off-the-shelf (COTS) FPGA boards.
This approach distinguishes ETA from many other radio telescope arrays using custom PCBs.
The design of complex, high speed boards reduces FPGA development time and cost advantages.
Although a single board integrating the receiver and beamforming functions is desirable, the effort required to design a custom PCB would be excessive even if no respins were required.
The low-cost evaluation boards meet the ETA system specifications remarkably well, and all interfacing is accomplished with standard cables and simple adapter boards containing only connectors and traces.

An additional benefit of COTS is straightforward upgrades to new FPGA families.
ETA2 can use a newer generation of FPGA evaluation boards: the ML410 \cite{ml410} and ML510 \cite{ml510} contain XC4VFX60 and XC5VFX130T FPGAs, respectively, while retaining the PC ATX form factor and personality module interfaces, and the Stratix-II DSP Development Kit \cite{dsp-2s60} contains a 2S60 device and preserves the MICTOR interface.
ML410 development boards are used in the LHC ATLAS beam conditions monitor \cite{LHC-ATLAS}.
Evaluation boards further extend the low cost and rapid development virtues of FPGAs, with the new boards costing roughly the same as their predecessors despite having greater logic capacity, higher performance, and expanded interfaces such as PCI Express and Serial ATA.
Common platforms such as BEE2 \cite{bee2-rcs} facilitate integration and IP reuse on large scale instruments developed by geographically separate teams.
Even if custom PCBs are justified, evaluation boards may still serve useful prototyping and validation roles.

We are developing an ML510 adapter board with eight Serial ATA connectors in order to have the FPGA write data directly to disks in a raw and interleaved manner.
This should improve sustained data recording rates by removing the variable latencies arising from a PC's operating system, file system, interrupts, multiplexed processors, caches and buses.
System cost, complexity and power are also reduced by having banks of inexpensive SATA disks replace server-class PCs containing data acquisition cards, SCSI disks and controllers.
After an acquisition, data can be retrieved through the ML510's Gigabit Ethernet interface.
Similar to the ML310-based cluster described in \cite{RCC-project}, inter-node communication can use the SATA ports rather than ETA's more expensive InfiniBand cables and connectors.
For streaming architectures, FPGAs with high-speed serial transceivers are often both necessary and sufficient for all computation, communication and data collection functions.

\begin{acks}
This work was supported by the National Science Foundation under  Grant No.\ AST-0504677, a SCHEV ETF grant, and by the Pisgah Astronomical Research Institute.
PARI personnel constructed the antenna masts and underground coaxial cable system.
Graduate student Vivek Venugopal performed data streaming tests on the Aurora links and data acquisition PCs.
We are grateful for early access to the ML310 board and the assistance of Vince Eck, Rick LeBlanc, Punit Kalra and Saeid Mousavi in Xilinx's Systems Engineering Group.
\end{acks}

\bibliographystyle{acmtrans}
\bibliography{eta_acm}

\begin{received}
\end{received}

\end{document}